\documentclass[12pt]{article}

\usepackage{amssymb,amsmath} %,amsthm,amsfonts}
\usepackage{epsfig} %,graphicx}
\newcommand{\beqs}{\begin{equation*}}
\newcommand{\beq}{\begin{equation}}

\newcommand{\eeqs}{\end{equation*}}
\newcommand{\eeq}{\end{equation}}

\newcommand{\beqas}{\begin{eqnarray*}}
\newcommand{\beqa}{\begin{eqnarray}}

\newcommand{\eeqas}{\end{eqnarray*}}
\newcommand{\eeqa}{\end{eqnarray}}

\newcommand{\eq}[2]{\begin{equation} #1 \label{#2} \end{equation}}

\newcommand{\al}{\alpha}
\newcommand{\be}{\beta}

\newcommand{\la}{\lambda}

\newcommand{\blist}{\begin{itemize}}

\newcommand{\elist}{\end{itemize}}

\providecommand{\href}[2]{#2}

\DeclareFontFamily{OT1}{rsfs}{}
\DeclareFontShape{OT1}{rsfs}{m}{n}{ <-7> rsfs5 <7-10> rsfs7 <10->rsfs10}{} 
\DeclareMathAlphabet{\mycal}{OT1}{rsfs}{m}{n}

\DeclareMathOperator{\extdm}{d}
\newcommand{\extd}{\extdm \!}

\newcommand{\dual}[1]{\tilde{#1}}

\newcommand{\Note}{note}

\begin{document}

%\begin{frontmatter}

%% use the thanksref command within \title, \author or \address for footnotes;
%% use the corauthref command within \author for corresponding author footnotes;
%% and the form \ead[url] for the home page:\title{}
%% \thanks[label1]{}
%% \author{Name\corauthref{cor1}\thanksref{label2}}
%% \thanks[label2]{}
%% \corauth[cor1]{}
%% \address{Address\thanksref{label3}}
%% \thanks[label3]{}

%\title{Duality in 2-dimensional dilaton gravity}


%\author{D.~Grumiller\corauthref{cor}}
%\ead{grumil@lns.mit.edu}
%\corauth[cor]{Corresponding author.}
%%
%and
%%
%\author{R.~Jackiw}
%\ead{jackiw@lns.mit.edu}
 
%\address{Center for Theoretical Physics,
%Massachusetts Institute of Technology,\\
%77 Massachusetts Ave.,
%Cambridge, MA  02139}

%\begin{abstract}

%We descry and discuss a duality in 2-dimensional dilaton gravity.

%\end{abstract}

%\begin{keyword}
%duality \sep 2-dimensional dilaton gravity 
%\PACS 04.60.Kz  \sep 04.70.Bw \sep 11.25.Tq
%\end{keyword}
%\end{frontmatter}

%\begin{document}

\begin{titlepage}

{\hfill MIT-CTP 3772}

{\hfill {\tt hep-th/0609197}}

\begin{center}

\vspace*{1.5truecm}

\textbf{\Large 
Duality in 2-dimensional dilaton gravity}

\vspace{10ex}

D.~Grumiller\footnote{e-mail: {\tt grumil@lns.mit.edu}} and R.~Jackiw\footnote{e-mail: {\tt jackiw@lns.mit.edu}}

  \vspace{7ex}

{\em Center for Theoretical Physics,
Massachusetts Institute of Technology,\\
77 Massachusetts Ave.,
Cambridge, MA  02139}

\end{center}

\vspace{14ex}

\begin{abstract}

We descry and discuss a duality in 2-dimensional dilaton gravity.

\end{abstract}

\end{titlepage}

%%% START OF THE TEXT %%%

\section{Introduction}

A large class of 2-dimensional (2D) gravity models is described by a general dilaton gravity action, which determines the line element in terms of a parameter in the action and an additional integration constant. We demonstrate that a well-defined transformation constructs another, inequivalent action, which belongs to the same class of models, leads to the same line element, but with the action parameter and integration constant interchanged. We call this transformation a duality: when carried out a second time, it reproduces the initial action.

In Section \ref{se:2}, we recall some results and present a specific example of the duality transformation. This example suggests the general procedure, which is described in Section \ref{se:3}. Section \ref{se:4} provides applications to various models and Section \ref{se:5} addresses coupling to matter.

\section{Recapitulation of some results}\label{se:2}

In a recent investigation \cite{Grumiller:2006ww} concerning geometry, we produced the 2D line element ($X\in[0,\pi]$)
\eq{
\extd s^2=2\extd u\extd X +\extd u^2\left(\la\cos{(X/2)+M}\right)\,,
}{eq:gj14}
or equivalently
\eq{
\extd s^2=\frac{1}{\cosh{(\dual{X}/2)}}\left[2\extd u\extd\dual{X}+\extd u^2\left(M\cosh{(\dual{X}/2)+\la}\right)\right]\,,
}{eq:gj18}
with  
\eq{
\tanh{(\dual{X}/2)}=\sin{(X/2)}\,.
}{eq:dual43}
We observed that \eqref{eq:gj14} or \eqref{eq:gj18} are solutions to the equations of motion that follow from the 2D gravity action (we use the same sign conventions as in \cite{Grumiller:2006ww})
\eq{
I=\frac{1}{8\pi^2}\int\extd^2x\sqrt{-g(x)}\left[X(x) R(x) + \frac{\la}{2} \sin{(X(x)/2)}\right]\,.
}{eq:gj15}
The parameter $M$ in \eqref{eq:gj14} and \eqref{eq:gj18} arises as an integration constant for the equations of motion implied by \eqref{eq:gj15}. These equations are solved in terms of two functions $X(x)$ and $u(x)$, which must allow expressing  uniquely $x^\al=x^\al(u,X)$, but otherwise are arbitrary. They are used as coordinates in \eqref{eq:gj14} or, after the redefinition \eqref{eq:dual43}, in \eqref{eq:gj18}. The parameter $\la$ appears in the action as a coupling constant with dimension inverse length squared, while the scalar (dilaton) field $X$  is dimensionless and $R$ is the Ricci scalar. The observation that we made is that the line element, directly in its representation \eqref{eq:gj18} or, after the redefinition \eqref{eq:dual43}, as \eqref{eq:gj14}, also arises as the solution to the equations of motion descending from the action
\begin{equation}
  \label{eq:gj20}
  \dual{I}=\frac{1}{8\pi^2}\int\extd^2x\sqrt{-g}\left[\dual{X}R+\frac12\tanh{(\dual{X}/2)}(\nabla \dual{X})^2-\frac{M}{4}\sinh{\dual{X}}\right]\,.
\end{equation}
While the two line elements \eqref{eq:gj14}, \eqref{eq:gj18} are related through the definition \eqref{eq:dual43}, $\dual{I}$ is not similarly related to $I$. Their functional forms differ and especially with \eqref{eq:gj20} $M$ is a parameter of the action and $\la$ occurs as an integration constant, which is opposite to the situation with \eqref{eq:gj15}.

In this {\Note} we show that the phe\-no\-me\-non observed in our example above is generally true for 2D dilaton gravity theories with the action 
\begin{equation}
  \label{eq:dual1}
  I=\frac{1}{8\pi^2}\int\extd^2x\sqrt{-g}\left[XR+U(X)(\nabla X)^2-\la V(X)\right]\,.
\end{equation}
The functions $U,V$ define the model, and many examples will be provided below. As before, the dilaton field $X$ is dimensionless, $\la$ is a parameter with dimension inverse length squared and $V$ is dimensionless. %The overall factor in front of the action will play no role in most of our considerations and has been chosen conveniently. 
We prove that for each action \eqref{eq:dual1} there is another one
which, although inequivalent to \eqref{eq:dual1}, leads to the same 2-parameter family of 2D geometries in a sense made precise below. We call the two actions duals of each other and we demonstrate how this works. 

In order to proceed we state some well-known results. For detailed explanations and references the review article \cite{Grumiller:2002nm} may be consulted.
The general solution for the line element derived from the action \eqref{eq:dual1} may be presented in Eddington-Finkelstein gauge as
\begin{equation}
  \label{eq:dual5}
  \extd s^2=e^Q \left[2\extd u \extd X + (\la w+M)\extd u^2\right]\,,
\end{equation}
with the constant of motion $M$ and the definitions
\eq{
Q^\prime(X):=-U(X)\,,\qquad w^\prime(X):=e^{Q(X)}V(X)\,,
}{eq:dualn1}
where prime means differentiation with respect to the argument. Two integration constants arise in the integrated versions of \eqref{eq:dualn1}. The first one (present in $Q$) is called ``scaling ambiguity'', the second one ``shift ambiguity''. We discuss later how to fix them appropriately; let us now just assume that they have been fixed in some way. Once the functions $Q$ and $w$ are known all other quantities may be derived without ambiguity.

Let us collect some properties of the solutions \eqref{eq:dual5}. Evidently, there is always (at least) one Killing vector $\partial_u$. The square of its norm is given by $e^Q(\la w+M)$, and therefore Killing horizons emerge for $X=X_h$, where $X_h$ is a solution of
\eq{
\la w(X_h)+M=0\,. 
}{eq:killinghorizons}
Like in the example \eqref{eq:gj14} the dilaton field is used as one of the coordinates in \eqref{eq:dual5}. This is possible globally, except at points where the Killing horizon bifurcates. 
The Ricci scalar is given by 
\eq{
R(X)=\la V^\prime(X) - 2\la V(X) U(X) - U^\prime(X) e^{-Q(X)} (M+\la w(X))\,, 
}{eq:Ricci}
and becomes $-U^\prime e^{-Q}\,M$ if the condition
\eq{
e^Q w = \rm constant
}{eq:MGS}
holds. This implies Minkowski spacetime for $M=0$, and hence a model with the property \eqref{eq:MGS} is called a ``Minkowski ground state model'' (MGS).

In addition to the family of line elements \eqref{eq:dual5} there are isolated solutions with constant dilaton vacuum (CDV) and maximally symmetric line element for each solution $X=X_{CDV}$ of the equation $V(X_{CDV})=0$. The corresponding Ricci scalar is given by
\eq{
R=\la V^\prime(X)|_{X=X_{CDV}}=\rm constant\,.
}{eq:CDV} 

\section{Duality in generic 2D dilaton gravity}\label{se:3}

To obtain the dual action we consider the definitions
\begin{equation}
  \label{eq:dual3}
  \extd\dual{X}:=\frac{\extd X}{w(X)}\,,\qquad e^{\dual{Q}(\dual{X})}\extd\dual{X}:= e^{Q(X)}\extd X\,,\qquad \dual{w}(\dual{X}):=\frac{1}{w(X)}\,.
\end{equation}
We assume that $w$ is strictly positive\footnote{If it is strictly negative similar considerations apply, but for sake of clarity we disregard this case. If $w$ has zeros but is bounded either from above or from below one may exploit the shift-ambiguity inherent to the definition of $w$ to give it a definite sign. If $w$ is unbounded from below and from above the definitions \eqref{eq:dual3} necessarily have singularities.} and therefore the definitions \eqref{eq:dual3} are well-defined (except for boundary values of $X$, typically either $X=0$ or $|X|=\infty$).
Differentiating \eqref{eq:dualn1} and inserting the definitions \eqref{eq:dual3} leads to the dual potentials
\begin{equation}
  \label{eq:dual23}
  \dual{U}(\dual{X})=w(X) U(X)-e^{Q(X)}V(X)\,,\qquad \dual{V}(\dual{X})=-\frac{V(X)}{w^2(X)}\,,
\end{equation}
which can be used to define a new action of the form 
\begin{equation}
  \label{eq:dual2}
  \dual{I}=\frac{1}{8\pi^2}\int\extd^2x\sqrt{-g}\left[\dual{X}R+\dual{U}(\dual{X})(\nabla \dual{X})^2 - M\dual{V}(\dual{X})\right]\,.
\end{equation}

The line element \eqref{eq:dual5}, following from the original action \eqref{eq:dual1}, is identical to the line element following from the dual action \eqref{eq:dual2} [with the dual potentials \eqref{eq:dual23}]. 
This can be shown as follows. We start with the action \eqref{eq:dual1} and the ensuing line element \eqref{eq:dual5}. Factoring $w$ in the latter allows presenting 
\eq{
\extd s^2= e^Qw\left[2\extd u \frac{\extd X}{w}+(\frac{M}{w}+\la)\extd u^2\right]\,.
}{eq:dual40}
Inserting the definitions \eqref{eq:dual3} yields 
 \begin{equation}
  \label{eq:dual15}
  \extd s^2=e^{\dual{Q}} \left[2\extd u\extd \dual{X} + (M \dual{w} + \la)\extd u^2\right]\,,
\end{equation}
But according to the general result \eqref{eq:dual5} this is the line element following from the dual action \eqref{eq:dual2} with dual potentials \eqref{eq:dual23}. The quantity $\la$ appears as a constant of motion in the dual formulation, so the roles of $M,\la$ are interchanged.
Evidently \eqref{eq:dual3} defines a line element-preserving diffeomorphism of $X$, with Jacobian $w$ and $e^Q$ transforming as a density.

We call the relation \eqref{eq:dual3} [together with \eqref{eq:dual23}] between the actions \eqref{eq:dual1} and \eqref{eq:dual2} a ``duality''. One reason for this nomenclature is that both of them generate the same set of geometries, as we have just shown, but with interchanged roles of parameter of the action and constant of motion. The other relevant observation is that the dual of the dual is always the original quantity (for $Q$, $w$, $U$, $V$ and $X$). %, $M$ and $\la$). 
So there is always a pair of actions related to each other in the way presented above, which justifies the use of the name ``duality''.

\subsection{Properties}

Here we list some general properties of the dual theories \eqref{eq:dual1} and \eqref{eq:dual2} [with \eqref{eq:dual3} and \eqref{eq:dual23}.]
\blist
\item The respective families of line elements are identical by construction; in particular, the number and types of Killing horizons are the same because every solution of \eqref{eq:killinghorizons} is also a solution of $M\dual{w}+\la=0$ and vice versa. Moreover, also the type of horizon remains the same: if it is extremal, i.e., if in addition to \eqref{eq:killinghorizons} also the relation $V(X_h)=0$ is fulfilled, then also the corresponding dual relation holds. Similarly, non-extremal horizons are also non-extremal horizons of the dual theory. Therefore, the respective global structures also coincide, except possibly in the asymptotic region and at the singularity, where $w$ and/or $e^Q$ may vanish or diverge. 
\item The respective line elements are not only the same, but also the number of CDVs where $V$ vanishes coincides for both theories, because $V=0$ implies with \eqref{eq:dual23} also $\dual{V}=0$ and vice versa. Moreover, if in addition to  $w>0$ also the inequalities $\la>0>M$ hold, then $(A)dS$ CDVs are dual to $(A)dS$ CDVs, because signs coincide of the respective Ricci scalars, given by \eqref{eq:CDV} and its dual version. Flat CDVs are always dual to flat CDVs.
\item MGS models, i.e., models with the property \eqref{eq:MGS}, are dual to models with $\dual{U}=0$.
\item Within the first order approach to 2D dilaton gravity the constant of motion $M$ has an interpretation as a Casimir function of a certain Poisson-sigma model \cite{Schaller:1994es}. If one replaces $\la$ in the action \eqref{eq:dual1} by a scalar field $B$ and adds a $BF$ term to the action, where $F$ is an Abelian field strength, then $B$ turns out to be a second Casimir function, the on-shell value of which is given by $\la$. The duality then acts by swapping these two Casimir functions. This mechanism can be extended to models which have even more Casimirs, for instance 2D dilaton gravity with additional gauge fields or with potentials $V$ containing coupling constants additional to $\la$.
\item Physically the duality exchanges the respective roles of reference mass [$\la$ in \eqref{eq:dual1} and $M$ in \eqref{eq:dual2}] and spacetime mass (as emerging from the constant of motion).
%-- analogously to the two possible ways an old fashioned balance may change its state: either the reference mass (corresponding to $\la$) is changed or the to-be-measured mass is changed (corresponding to $M$). For an observer looking at the balance (but not at the masses) both situations are indistinguishable.
\elist 
It should be noted that our duality, which relates different actions but leaves invariant the space of solutions for the metric, is quite different from the target space duality in \cite{Giveon:1991sy,Cadoni:1994vs}, which acts on the space of solutions and leaves invariant the gauge fixed action.

\subsection{Fixing the ambiguities}

The functions $w$ and $e^Q$ are defined \eqref{eq:dualn1} up to two integration constants only, corresponding to a shift ambiguity $w\to w+\be$ and a scale ambiguity $e^Q\to\al e^Q$ and $w\to\al w$, where $\al,\be$ are some real numbers. The shift ambiguity can be fixed for a large class of models as follows: The duality \eqref{eq:dual3} is meaningful only if $w$ does not have any zeros. If it has zeros but is bounded from above or from below (in the range of definition of the dilaton), one may exploit the shift ambiguity to eliminate all zeros. If one does this in such a way that the only zero of $w$ lies at either of the boundaries (typically $X=0$ or $|X|=\infty$) then the shift ambiguity is fixed uniquely. We shall exhibit how this works for concrete examples below. Physically, this ambiguity corresponds to a choice of the ground state solution, $M=0$. Often there is a preferred choice, such as a maximally symmetric spacetime, but if there is none then one just has to choose any particular solution as the ground state. 
The scaling ambiguity is harmless for classical considerations and may be absorbed by a rescaling of the coordinate $u$ together with an appropriate rescaling of the mass $M$. 
Additionally, the representation of the potential $V$ in the action \eqref{eq:dual1} is ambiguous because one may multiply $\la$ by some dimensionless constant and divide $V$ through the same constant. This ambiguity is not essential, because it can be absorbed into the scaling ambiguity discussed above.

Some definitions involve sign ambiguities. Without loss of generality we require $\la>0$ and, as mentioned before, for sake of definiteness also $w>0$. For solutions which have a Killing horizon from \eqref{eq:killinghorizons} we deduce $M<0$. Thus {\em minus} $M$ is directly related to the physical mass of black hole (BH) solutions. In the dual formulation the situation is reversed, i.e., {\em plus} $\la$ is directly related to the physical mass. As an illustration we consider the model $w=e^{-Q}=\sqrt{X}$. Introducing $\extd r=e^Q\extd X$ and inserting the functions $Q$, $w$ into the line element \eqref{eq:dual5} yields
$\extd s^2=2\extd u \extd r +(\la +2M/r)\extd u^2$,
which is the 2D part of the Schwarzschild (S) BH in Eddington-Finkelstein gauge. However, $r$ is dimensionless, $u$ has dimension length squared and $\la, M$ have dimension inverse length squared. Therefore, we redefine $\hat{u}:=\sqrt{\la}u$, $\hat{r}:=r/\sqrt{\la}$ and obtain
\eq{
\extd s^2=2\extd\hat{u} \extd\hat{r} +\left(1 - \frac{2M_{ADM}}{\hat{r}}\right)\extd\hat{u}^2\,,
}{eq:ADM2}
with  $M_{ADM} = -M\la^{-3/2}$ in units of length. For fixed positive $\la$ the constant of motion $-M$ determines the ADM mass, while in the dual formulation for fixed negative $M$ the constant of motion $\la^{-3/2}$ determines the ADM mass.

\section{Examples and applications}\label{se:4}

As a demonstration let us now apply the definitions and results from the previous Section to the example presented before, \eqref{eq:gj14}-\eqref{eq:gj20}, with particular emphasis on various ambiguities. Starting point is the action \eqref{eq:gj15}. The function $U$ vanishes in that case, while $V\propto \sin{(X/2)}$, and  $w\propto\cos{(X/2)}+c$. We may fix the shift ambiguity by setting $c=0$ so that $w>0$ for $X\in[0,\pi)$ and $w\to 0$ for $X\to\pi$. The scaling ambiguity is fixed by identifying $\la$ in \eqref{eq:gj15} with $\la$ in \eqref{eq:dual1} in order to reproduce \eqref{eq:dual43} with the same numerical factors. This gives $V=-\frac12\sin{(X/2)}$ and $w=\cos{(X/2)}$. The formulas \eqref{eq:dual3}, \eqref{eq:dual23} lead to $\dual{U}=\frac12\tanh{(\dual{X}/2)}$, $\dual{V}=\frac14\sinh{\dual{X}}$, $\dual{Q}=-\ln{\cosh{(\dual{X}/2)}}$ and $\dual{w}=\cosh{(\dual{X}/2)}$, which after insertion into \eqref{eq:dual2} correctly reproduces \eqref{eq:gj20}.

Next we apply the general procedure to several models. Among the best-known 2D dilaton gravity models are the S BH (spherically reduced to 2D), the Jackiw-Teitelboim (JT) model \cite{Jackiw:1984} %,Teitelboim:1984} 
and the Witten (W) BH \cite{Witten:1991yr}. %,Mandal:1991tz,Elitzur:1991cb,Callan:1992rs}. 
It was realized in \cite{Katanaev:1997ni} that all of them can be summarized in a 2-parameter family of 2D dilaton gravity models of the form
\begin{equation}
  \label{eq:dual6}
  e^Q=X^{-a}\,,\quad w=X^{b+1}\quad\rightarrow\quad U=\frac{a}{X}\,,\quad V=(b+1)X^{a+b}\,.
\end{equation}
The shift ambiguity inherent to $w$ has been fixed by requiring $w(0)=0$ for $b>-1$ and $w(\infty)=0$ for $b<-1$ (for $b=-1$ there is no preferred way to fix this ambiguity, 
so we choose $w=1$). The scale ambiguity has been fixed conveniently. The line element \eqref{eq:dual5} may be re-parameterized as ($X^{-a}\extd X=\extd r$)
\eq{
\extd s^2=2\extd u \extd r + \left(\la X^{b+1-a}(r)+MX^{-a}(r)\right)\extd u^2\,.
}{eq:ab1}
The Ricci scalar \eqref{eq:Ricci} reads 
\eq{
R=\la b (b+1-a) X^{a+b-1} + M a X^{a-2}\,.
}{eq:ab2}
Notably $R$ is constant for all $\la, M$ if and only if $(a-1)^2=1$ and $b(b^2-1)=0$. When $a-b=1$ the MGS property \eqref{eq:MGS} holds. If $a+b=1$ the ground state solution ($M=0$) is $(A)dS$. For $a=0$ $R$ becomes independent from $M$.  Models with $a\neq 1$, $b=0$ are called Rindler ground state models, because $R$ vanishes and the Killing norm is linear in the coordinate $r$ for $M=0$. We shall assume $b\neq 0$ and discuss this special case separately at the end. Applying our rules \eqref{eq:dual3} to the models given by \eqref{eq:dual6} yields 
\eq{
e^{\dual{Q}}= (-b\dual{X})^{-1+(a-1)/b}\,,\qquad \dual{w}=(-b\dual{X})^{1+1/b}\,.
}{eq:dual55}
After the field redefinition $-b\dual{X}\to\dual{X}$ the functions $\dual{Q}$ and $\dual{w}$ are again in the form \eqref{eq:dual6}, so our duality maps one model of the $ab$ family to another one of the same family with new parameters $\dual{a}$ and $\dual{b}$ given by
\begin{equation}
  \label{eq:dual7}
  \dual{a} = 1-\frac{a-1}{b}\,,\qquad \dual{b}=\frac 1b \,.
\end{equation}
The fixed points under duality transformations are $b=a=1$ and $b=-1$, $a$ arbitrary. 
% Sometimes models which have the same $b$ but different values of $a$ are considered to be equivalent, as the difference between the corresponding line elements \eqref{eq:dual5} is just a globally regular conformal factor (possibly apart from boundary points where $e^Q$ may vanish or diverge). In that case the whole class of theories conformally related to the JT model, $b=1$ and $a$ arbitrary, is a fixed point. 

\begin{figure}
\centering
\epsfig{file=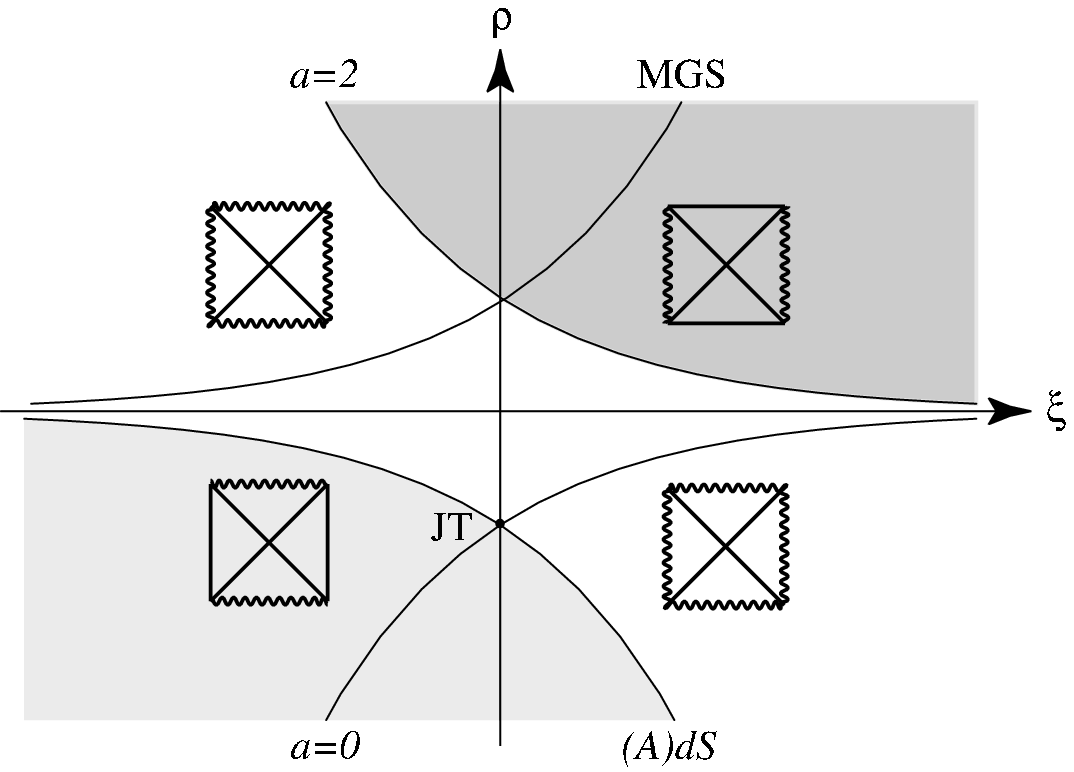,width=0.8\linewidth}
\caption{$b>0$}
\label{fig:1} 
\end{figure}

\begin{figure}
\centering
\epsfig{file=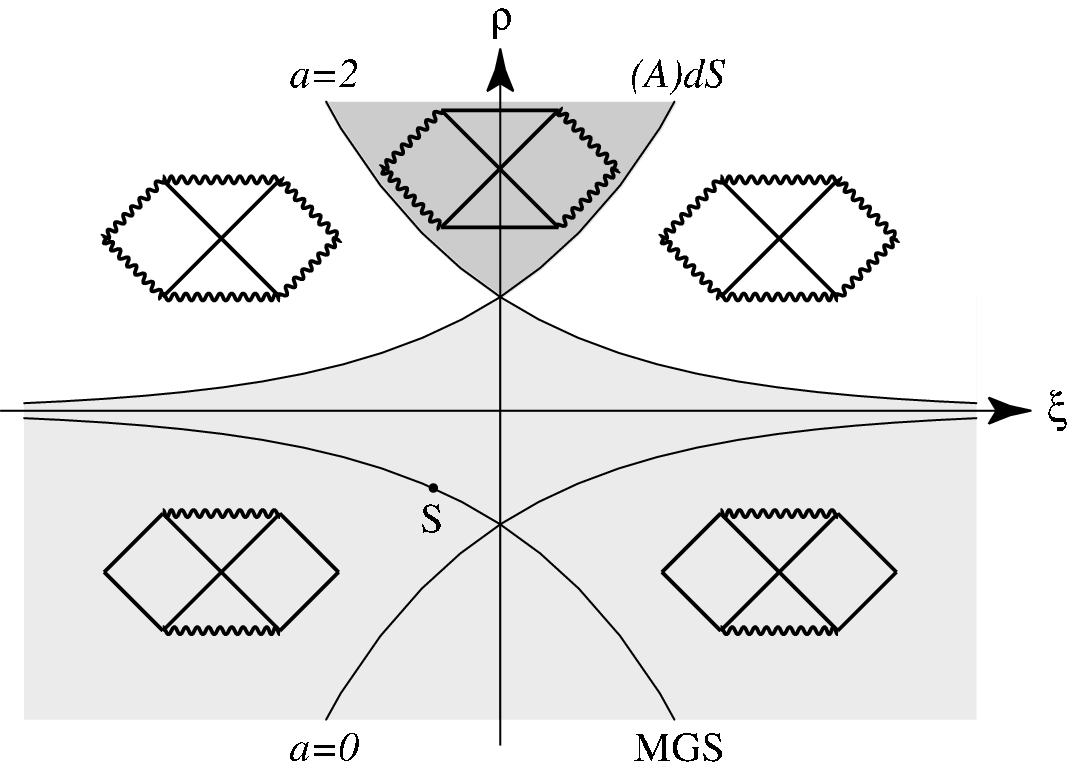,width=0.8\linewidth}
\caption{$b<0$}
\label{fig:2} 
\end{figure}

It should be mentioned that the combination
\eq{
\rho:=\frac{(a-1)}{\sqrt{|b|}}={\rm sign\,}{(-\dual{b})} \frac{(\dual{a}-1)}{\sqrt{|\dual{b}|}}
}{eq:rho}
is invariant under the duality for $b<0$ and goes to $\dual{\rho}=-\rho$ for $b>0$. This leads to a useful representation of the ``phase space'' of Carter-Penrose diagrams. In Fig.~6 of \cite{Katanaev:1997ni} that phase space is depicted as function of $a$ and $b$. 
We present the same graph as a function of $\rho$ and $\xi=\ln{\sqrt{|b|}}$, discriminating between positive $b$ (Fig.~\ref{fig:1}) and negative $b$ (Fig.~\ref{fig:2}). In the former case duality acts by reflection at the origin, in the latter case duality acts by reflection at the $\rho$ axis. So in Fig.~\ref{fig:1} the white region is mapped onto itself (with the origin as fixed point), whereas the light and dark gray regions are mapped onto each other by duality. By contrast, in Fig.~\ref{fig:2} each of the three differently shaded regions is mapped onto itself (with the $\rho$ axis as line of fixed points). Therefore, in Fig.~\ref{fig:1} the singularities and asymptotic regions are exchanged by the duality, while in Fig.~\ref{fig:2} they remain the same. The four exponential curves in both graphs correspond to the line of MGS models ($a=1+b$), models which have an $(A)dS$ ground state ($a=1-b$), models with no kinetic term for the dilaton ($a=0$) and models with $a=2$. At the intersection points of these curves lie models which have maximally symmetric spacetimes for any value of $\la, M$. The JT model appears in Fig.~\ref{fig:1}, the S BH in Fig.~\ref{fig:2}. The W BH emerges as an asymptotic limit ($\xi\to-\infty$ on the $\xi$-axis) in both graphs.

In particular, for spherically reduced models from $D$ dimensions we have $b=-1/(D-2)$ and $a=(D-3)/(D-2)$. Since $X$ has a higher dimensional interpretation as surface area it is fair to ask for a physical interpretation of $\dual{X}$. It may be checked easily that $\dual{X}\propto X^{1/(D-2)}$ is the surface radius. The dual model is conformally related\footnote{By ``conformally related'' we mean that the difference between the corresponding line elements is a conformal factor which is regular globally, except possibly at boundary points where $e^Q$ may vanish or diverge.} to spherically reduced gravity from $\dual{D}$ dimensions, with
$\dual{D}=(2D-3)/(D-2)$. %\,.
Obviously, only $D=3$ is self-dual, while the S BH ($D=4$) is related to $\dual{D}=5/2$. We mention that an alternative representation of the dual action is obtained by eliminating the dual dilaton field by means of its equation of motion, $\dual{X}\propto R^{1/(1-D)}$:
\eq{
\dual{I} = N_D \int\extd^2x\sqrt{-g} \,R^{1+1/(1-D)}\,,
}{eq:dual83}
where $N_D$ is a constant following from the normalization chosen in \eqref{eq:dual2}. It grows with $D$ for $D\gg 1$. This class of theories has been studied in \cite{Schmidt:1991ws}.

The JT model ($a=0$, $b=1$) is very special as it is not only dual, but also conformally related to the model $\dual{a}=2$, $\dual{b}=1$. 

Finally we discuss the models with $b=0$. We assume first $a\neq 1$, so that the ground state solution is Rindler spacetime. The dual dilaton is given by $\dual{X}=\ln{X}$. Therefore, the dual model does not belong to the $ab$-family; rather it belongs to the class of Liouville gravities \cite{Nakayama:2004vk,Jackiw:2005su}: $\dual{Q}=(1-a)\dual{X}$ and $\dual{w}=e^{-\dual{X}}$. In particular, for $a=0$ the CGHS model emerges (cf.~the last Ref.~\cite{Witten:1991yr}). Its dual is given by the ``almost Weyl invariant'' Liouville model \cite{Jackiw:2005su}.
In \cite{Bergamin:2004pn} it was observed that the Ricci scalar obtained from Liouville gravity is independent from the constant of motion. This feature is simply a consequence of the duality of Liouville gravity to Rindler ground state models. The W BH ($a=1$, $b=0$) has the MGS property \eqref{eq:MGS}. Its dual is given by $\dual{Q}=0$ and $\dual{w}=e^{-\dual{X}}$, thus leading to the dual action
\eq{
\dual{I}_{WBH}=\frac{1}{8\pi^2}\int\extd^2x\sqrt{-g}\left(\dual{X}R+Me^{-\dual{X}}\right)\,.
}{eq:dualWBH1}
Elimination of the dual dilaton field by means of its equation of motion, 
$\dual{X}=-\ln{(R/M)}$, %\,,
and re-insertion into \eqref{eq:dualWBH1} allows to represent the dual W BH action (up to Einstein-Hilbert terms) as
\eq{
\dual{I}_{WBH}=-\frac{1}{8\pi^2}\int\extd^2x\sqrt{-g}\,R\ln{|R|}\,.
}{eq:dualWBH2}
This dual action for the W BH was presented for the first time in \cite{Frolov:1992xx} and it arises also as the $D\to\infty$ limit of \eqref{eq:dual83}, concurrent with the fact that the $D\to\infty$ limit of spherically reduced gravity yields the W BH. This limiting procedure resembles the one discussed in \cite{Jackiw:2005su}.

\section{Outlook}\label{se:5}

The duality \eqref{eq:dual3} [together with \eqref{eq:dual23}] between the actions \eqref{eq:dual1} and \eqref{eq:dual2} leaves intact the line element \eqref{eq:dual5} but changes the dilaton field. This has some consequences for coupling to matter as well as for thermodynamical and semi-classical considerations, which we shall outline briefly. Generally any phe\-no\-me\-non that is based upon a fixed background geometry and that is not sensitive to the dilaton field will be invariant under the duality, but even quantities that are sensitive to the dilaton field (like quasi-normal modes of a scalar field) may be duality-invariant.

An example of a duality-invariant observable is the Hawking temperature, as derived either naively from surface gravity or from the Hawking flux of a Klein-Gordon field propagating on the (fixed) BH background. 
An example of a duality-non-invariant observable is the Bekenstein-Hawking entropy, which is proportional to the dilaton field \cite{Gegenberg:1995pv} and thus changes under the duality. 

It would be interesting to find observables that are duality-invariant in the full dynamical and self-consistent system of geometry plus matter, i.e., not based upon some fixed background approximation.

\section*{Acknowledgments}

DG is grateful to Max Kreuzer, Wolfgang Kummer, Alessandro Torrielli, Dimitri Vassilevich for discussions and to Ralf Lehnert for help with the Figures. We thank Mariano Cadoni for calling our attention to \cite{Cadoni:1994vs}.

This work is supported in part by funds provided by the U.S. Department of Energy (DOE) under the cooperative research agreement DEFG02-05ER41360.
DG has been supported by the Marie Curie Fellowship MC-OIF 021421 of the European Commission under the Sixth EU Framework Programme for Research and Technological Development (FP6).

%\bibliographystyle{fullsort}
%\bibliographystyle{h-elsevier2.bst}
%\bibliography{review}

\providecommand{\href}[2]{#2}\begingroup\raggedright\endgroup

\end{document}